\documentclass[namedreferences]{SolarPhysics}
\usepackage[optionalrh]{spr-sola-addons} 
\usepackage[pdftex]{graphicx} 
\usepackage{color}           
\usepackage{url}             

\newcommand{\rsun}{$R_\odot$}
\newcommand{\degree}{$^{\circ}$}

\begin{document}

\begin{article}

\begin{opening}

\title{CME Propagation Characteristics from Radio Observations}

\author{S.~\surname{Pohjolainen}$^{1}$\sep
        L.~\surname{van Driel-Gesztelyi}$^{2,3,4}$\sep
        J.L.~\surname{Culhane}$^{2}$\sep
        P.K.~\surname{Manoharan}$^{5}$\sep       
        H.A.~\surname{Elliott}$^{6}$}  

\runningauthor{Pohjolainen et al.}
\runningtitle{CME Propagation Characteristics}

\institute{$^{1}$ Tuorla Observatory/Department of Physics, University of 
                  Turku, V\"ais\"al\"antie 20, 21500 Piikki\"o, Finland 
                  email: \url{silpoh@utu.fi}\\
           $^{2}$ Mullard Space Science Laboratory, University College London, 
                  Holmbury St. Mary, Dorking, Surrey, RH5 6NT, UK\\
           $^{3}$ Observatoire de Paris, LESIA, FRE 2461 (CNRS),
                  92195 Meudon Principal Cedex, France\\
           $^{4}$ Konkoly Observatory of the Hungarian Academy of Sciences, 
                  Budapest, Hungary\\
           $^{5}$ Radio Astronomy Centre, Tata Institute of Fundamental 
                  Research, India\\
           $^{6}$ Southwest Research Institute, San Antonio, Texas, USA\\
             }

\date{Solar Physics, accepted July 2007}

\begin{abstract}
\sloppy{We explore the relationship among three coronal mass ejections
(CMEs), observed on 28 October 2003, 7 November 2004, and 20 January
2005, the type II burst-associated shock waves in the corona and solar 
wind, as well as the arrival of their related shock waves and magnetic 
clouds at 1 AU}. Using six different coronal/interplanetary density models, we
calculate the speeds of shocks from the frequency drifts observed in
metric and decametric radio wave data.  We compare these speeds with 
the velocity of the CMEs as observed in the plane-of-the-sky white-light 
observations and calculated with a cone model for the 7 November 2004 event. 
We then follow the propagation of the ejecta using Interplanetary 
Scintillation (IPS) measurements, which were available for the 
\mbox{7 November 2004} and 20 January 2005 events. 
Finally, we calculate the travel time of the interplanetary (IP) 
shocks between the Sun and Earth and discuss the velocities obtained 
from the different data. This study highlights the difficulties in 
making velocity estimates that cover the full CME propagation time.
\end{abstract}
\keywords{Coronal Mass Ejections, Initiation and Propagation, Interplanetary;
Plasma Physics; Radio Bursts, Dynamic Spectrum, Meter-Wavelengths
and \\ Longer, Type II; Radio Scintillation} 
\end{opening}

\section{Introduction}

In order to deduce reliable values of arrival time at Earth for
CME-related ejecta, it is first necessary to determine the CME
departure speed from the Sun where the phases of slow rise,
acceleration, constant velocity, and deceleration all need to be
considered. The problem is made more difficult by the fact that
Earth-directed events are so-called halo or partial-halo CMEs that
have a quasi-symmetric appearance around the solar disc when viewed in
white light. Thus it is more difficult to obtain true earthward
velocities at the departure times of these events.

A range of tools exists for the estimation of these speeds that are
mostly applicable from the lower corona to distances up to $r \sim$ 
10 \rsun. These include the observation of radio bursts in the 
dynamic spectra (principally type II bursts) and the analysis of 
white-light images obtained with coronagraphs. For values 
\mbox{$r >$ 50 \rsun}, when the faster CMEs are usually decelerating, 
the Interplanetary Scintillation (IPS) technique can be used. 
While it is possible to track the expansion of EUV or X-ray emitting 
plasmas by using appropriate imagers or by directly measuring
velocities spectroscopically, these methods have been applied to CMEs 
in just a few isolated cases. While we expect a greater use of both 
imaging and spectroscopy in the future, through data obtained by the 
{\it Hinode} and STEREO missions, neither of these latter methods has 
been employed for the events discussed here.

Since it is important to understand the strengths and limitations of
these methods before applying them to CME speed estimation, it is
necessary to briefly describe the underlying physics. We therefore
present a short survey in the next section that outlines the basis for
use of radio-burst dynamic spectra and white-light images for speed
measurement. For $r >$ 50 \rsun, use of the IPS technique is most
appropriate, and so we outline the basic features of that method. 
We then describe the application of these techniques to the events that
had been selected for the Sun-Earth Connection Workshop (in this Topical 
Issue), namely those of 28 October 2003, 7 November 2004, and 
20 January 2005, and the results obtained from the analysis. We 
conclude with a general discussion of all three halo CMEs and the 
uncertainties in speed estimation.

\section{Speed Measuring Techniques and Underlying Mechanisms}

Frontside ``halo'' CMEs are directed towards the Earth and appear as
diffuse clouds that surround the solar disk in white-light images.  
Some of them can be very geoeffective (Kim {\it et al.}, \citeyear{kim05}; 
Howard and Tappin, \citeyear{howard05}). 
Halo CMEs are often very fast, the measured apparent speeds are
957 km s$^{-1}$ on average \cite{yashiro04}, which is about 
twice the mean CME speed. Speed estimation can, however,
be difficult \cite{michalek03}, as CMEs are observed in scattered 
white light on the plane of the sky, which is biased by projection 
effects, sideways expansion of the structure, and its propagation 
towards us. Using both white-light and radio observations, 
the ``true'' CME speeds have in some studies been 
estimated to be higher than the measured plane-of-the-sky 
speeds \cite{reiner03}.

Solar type II bursts are slow-drift bursts visible in dynamic radio 
spectra, and they are generally attributed to shock-accelerated
electrons (Wild and Smerd, \citeyear{wild72}; Nelson and Melrose,  
\citeyear{nelson85}). 
A distinction can be made between coronal type II bursts observed in 
the decimetric--metric wavelength range and interplanetary (IP) type 
II bursts observed at decametric--hectometric (DH) wavelenghts. 
The driving agent of type II bursts has also been under discussion: 
shock acceleration can be created by blast waves ({\it e.g.}, a pressure pulse 
without mass motions driving the wave) or piston-driven shocks (mass
propagating at super-Alfv\'enic speed). There has been discussion on
whether coronal shocks can survive into the IP space ({\it e.g.}, 
Mann {\it et al.}, \citeyear{mann03}; Knock and Cairns, \citeyear{knock05}).   
IP type II bursts are usually ascribed to bow shocks driven ahead of 
a CME \cite{kahler92}, while coronal type II bursts have shown better 
temporal and spatial correlation with flare waves and ejecta ({\it e.g.}, 
Klassen, Pohjolainen, and Klein, \citeyear{klassen03}; 
Cane and Erickson, \citeyear{cane05}; 
Cliver {\it et al.}, \citeyear{cliver05}, and references therein).

The early work on large-scale propagating transients was based on 
IPS measurements of a large number of radio sources at low
frequencies, applicable to relatively large distances from the Sun 
(see, {\it e.g.}, Hewish, Tappin, and Gapper, \citeyear{hewish85}). 
Present-day measurements at higher frequencies provide CMEs' size, 
speed, turbulence level, and mass also closer to the Sun, at 
distances $r >$ 50 \rsun \ ({\it e.g.}, Manoharan {\it et al.}, 
\citeyear{mano95}; Tokumaru {\it et al.}, \citeyear{tokumaru03}).    
The regular monitoring of IPS on a given radio source over several
days provides variations of the solar wind speed and density turbulence 
at a large range of heliocentric distances from the Sun. IPS 
observations of a grid of large number of radio sources on consecutive 
days can provide three-dimensional images of the heliosphere at 
different radii.

\subsection{CME OBSERVATIONS IN WHITE LIGHT}

Since the CME structures are seen in the plane of the sky, projection
effects must be considered. 
Several methods for calculating radial velocities have recently been 
proposed by, {\it e.g.}, Leblanc {\it et al.} (\citeyear{leblanc01}), 
Micha\l ek, Gopalswamy, and Yashiro (\citeyear{michalek03}, symmetric cone
model),  Schwenn {\it et al.} (\citeyear{schwenn05}, lateral expansion speed),
and Micha\l ek (\citeyear{michalek06}, asymmetric cone model). 
We selected the symmetric cone model by Micha\l ek, Gopalswamy, and 
Yashiro (\citeyear{michalek03}) for estimating the "true" speeds 
assuming that for a halo CME, propagation is with constant velocity 
and angular width, and the bulk velocity is directed radially and isotropic. 
If the launch location of the CME is slightly shifted along the cone 
symmetry axis by a distance $r$ with respect to Sun centre, 
the initial and final appearance of the halo CME will be at the
opposite limbs on this axis. Using measured plane-of-the-sky
velocities at diametrically opposite limbs and the time difference, 
$\Delta$$T$ between first and last appearance, the cone model may be used 
to determine $r$ along with {\it i}) $\gamma$, the angle between the 
symmetry axis and the plane-of-the-sky, {\it ii}) $\alpha$, the
opening angle of the cone, and {\it iii}) $V$, the radial CME velocity.

In view of the difficulty in estimating $\Delta$$T$, given the relatively poor
cadence of LASCO measurements, and since the active-region launch sites
were identified for the events, we established $r$ from MDI
magnetogram and EIT image data, thus allowing an estimate of  $\gamma$ from the
relation $\cos \gamma$ = $r/$\rsun. Values of $\alpha$ and $V$ were then 
deduced from the Equations (3) and (4) of Micha\l ek, Gopalswamy, and
Yashiro (\citeyear{michalek03}).
In fact the application of this technique was possible for only one of 
our events, that of 7 November 2004. For the event of 28 October 2003, only 
three useable LASCO C3 frames were available and the velocities at opposite 
limbs were essentially the same, making it impossible to apply the cone model. 
In the case of the event of 20 January 2005, only one useful LASCO frame 
was available.

\subsection{RADIO OBSERVATIONS: FROM PLASMA FREQUENCY TO SHOCK VELOCITY}

The general idea behind type II bursts is that they are created by 
a propagating shock: Langmuir waves are excited by electron beams produced 
in this shock and the waves are then converted into escaping radio 
waves (Melrose \citeyear{melrose80}; Cairns {\it et al.} \citeyear{cairns03},  
and references therein). 
The emission mechanism is plasma emission near the fundamental and 
second-harmonic frequencies. 

The plasma frequency $f_p$ (Hz) at the fundamental is directly related 
to the electron density $n_e$ (cm$^{-3}$) by 
 \begin{equation}
f_p=9000\sqrt{n_e} \  ,
\end{equation}
and a frequency drift towards the lower frequencies shows that the 
electron density is falling. This change is usually attributed to 
the burst driver moving in the solar atmosphere towards lower 
densities and larger heights. Type II bursts are therefore a valuable 
tool in determining burst (shock) velocities.  

Identifying the fundamental emission lane in dynamic radio spectra can
be tricky (see, {\it e.g.}, Robinson, \citeyear{robinson85a}). Firstly  
not all type II bursts show emission at the fundamental and second 
harmonic (second harmonic should be near 2$f_p$), and due 
to the limited observing frequencies it is 
possible that one or the other lane is not visible in the spectrum.
The DH type II bursts are characterised by weaker emission than 
their metric counterparts, and the emission lanes are often split 
into a series of patches. In metric bursts the harmonic band can 
be stronger than the fundamental, but in DH bursts it is usually
the opposite (see Vr\u{s}nak {\it et al.}, \citeyear{vrsnak01} and references 
therein). Both the fundamental and second harmonic can show 
band-splitting (Vr\u{s}nak {\it et al.}, \citeyear{vrsnak01}; 
\citeyear{vrsnak02}; Vr\u{s}nak, Magdaleni\u{c}, and Zlobec,  
\citeyear{vrsnak04}). 
Sometimes rapidly drifting emission stripes 
called ``herringbones'' can be observed over a type II burst 
``backbone'' (see, {\it e.g.}, Cairns and Robinson, \citeyear{cairns87}; 
Mann and Klassen, \citeyear{mann05}). Ground-based observations 
at decimetric-metric wavelengths are also easily affected by radio
interference.  

\begin{figure}
\begin{center}
\includegraphics[width=0.9\textwidth]{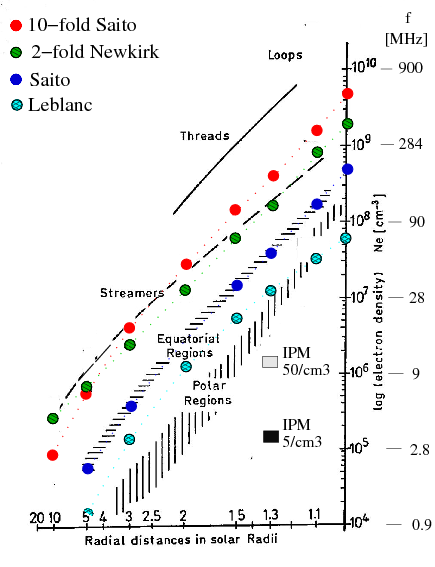}  
\caption{Electron density in different parts of the corona from eclipse
photometry (image courtesy of S. Koutchmy, see also Koutchmy, 1994).
Dark blue circles: coronal densities according to the Saito (1970) model,  
red circles: densities according to the ten-fold Saito model, green 
circles: densities according to the two-fold Newkirk (1961) model,
light blue circles: densities according to the Leblanc, Dulk, and Bougeret 
(1998) model. IPM-labeled boxes indicate electron densities at 
1.3 {\rsun} \ according to the IP density model, with near-Earth 
electron densities of 5 cm$^{-3}$ (solar minimum) and 50 cm$^{-3}$ 
(solar maximum).}  
\label{fig1}      
\end{center}
\end{figure}

After calculating the electron density from the plasma frequency 
at the fundamental, the next step is to find a corresponding height 
for the emitting source. This is done with the help of atmospheric 
density models. Which model to use is a well-known problem, see, {\it e.g.}, 
Robinson and Stewart (\citeyear{robinson85b}) and references therein.
Figure~\ref{fig1} shows how density depends strongly 
on coronal conditions: it is  important to know if the disturbance 
is propagating in a less-dense equatorial region, inside a dense 
streamer region, or in even denser coronal-loop structures. Also the 
turbulent afterflows of a previous CME can affect the densities.
The most widely used density models are by Newkirk (\citeyear{newkirk61}) 
and Saito (\citeyear{saito70}). In the Newkirk model, the electron number 
densities stay high at large distances from the Sun since the model 
is a hydrostatic one, see Figure~\ref{fig1}. 

The observed density values near 1 AU are much less than in the corona 
(see, {\it e.g.}, Mann {\it et al.} \citeyear{mann99}; \citeyear{mann03}), and 
the large density decreases are mainly due to solar-wind densities.   
In interplanetary (IP) space a commonly-used approximation 
$n_e \sim R^{-2}$ leads to $f_p \sim R^{-1}$, and in this case  
the plasma density $n_e$ (cm$^{-3}$) scales as 
   \begin{equation}
    n_e=\frac{n_0}{R_d^2} \  ,
   \end{equation}
where $n_0$ (cm$^{-3}$) is the plasma density near Earth at 1 AU, and 
$R_d$ (in AU) is the distance from the Sun ({\it e.g.}, Reiner {\it et al.},  
\citeyear{reiner01}). The value of $ n_0$ ranges from $\sim$5 cm$^{-3}$ 
around solar minimum to ten times this value at solar maximum. 

The model of Saito (\citeyear{saito70}) departs significantly in 
IP space from the $R^{-2}$ scaling. A revised model by Saito, Poland, 
and Munro (\citeyear{saito77}) is similar to that of Saito 
(\citeyear{saito70}) in the $R$ = 1.2\,--\,10 {\rsun} range, but 
the densities get closer to the $n_e \sim R^{-2}$ relation at larger 
distances. 
Also the model by Leblanc, Dulk, and Bougeret (\citeyear{leblanc98}) gives 
too low densities for the active region corona (the authors recommend
not to use it below $\sim$1.2 {\rsun}), but it is similar to the
IP model at distances $>$ 30 {\rsun}. A ``hybrid-model" by 
Vr\u{s}nak, Magdaleni\u{c}, and Zlobec (\citeyear{vrsnak04}) is a 
mixture of a five-fold Saito and the Leblanc, Dulk, and Bougeret
models, with small modifications, and it can be 
used for connecting bursts in the corona and the IP space. Both give 
an electron density of 7 cm$^{-3}$ near 1 AU, which is a commonly 
observed value during activity minimum. 

\begin{table}[!hb]
\caption{From plasma frequency to heliocentric burst driver 
height (\rsun) with different atmospheric density models}
\label{table1}
\begin{tabular}{ccccccccc}
\hline
\smallskip
$f_p$ & $\lambda$ & $n_e$       & $h$  & $h$ & $h$       & $h$ & $h$ & $h$ \\
      &           &             &      &     & 2$\times$ & 10$\times$ & & \\
(MHz) & (m)       & (cm$^{-3}$) & Saito& Hybrid   & Newkirk & Saito & 
Leblanc$^{\bf a}$ & IP$^{\bf a}$\\
\hline
500 & ~0.6 & 3.1$\times$10$^9$ &-    & -     &-      & ~1.03  &-     &- \\
400 & ~0.7 & 2.0$\times$10$^9$ &-    & -     &-      & ~1.08  &-     &- \\ 
300 & ~1.0 & 1.1$\times$10$^9$ &-    & ~1.04 &~1.05  & ~1.14  & -    &- \\ 
200 & ~1.5 & 4.9$\times$10$^8$ &-    & ~1.12 &~1.15  & ~1.26  & -    &- \\
100 & ~3.0 & 1.2$\times$10$^8$ &1.13 & ~1.30 &~1.37  & ~1.56  &-     &- \\
70  & ~4.3 & 6.0$\times$10$^7$ &1.23 & ~1.45 &~1.51  & ~1.76  &-     &- \\
50  & ~6.0 & 3.1$\times$10$^7$ &1.34 & ~1.64 &~1.68  & ~1.99  & 1.10 &- \\
30  & 10.0 & 1.1$\times$10$^7$ &1.58 & ~2.01 &~2.04  & ~2.40  & 1.31 &- \\
14  & 21.4 & 2.4$\times$10$^6$ &2.07 & ~2.78 &~2.96  & ~3.33  & 1.71 &- \\
12  & 25.0 & 1.8$\times$10$^6$ &2.19 & ~2.96 &~3.24  & ~3.57  & 1.80 &- \\
10  & 30.0 & 1.2$\times$10$^6$ &2.36 & ~3.24 &~3.74  & ~3.97  & 1.93 &-\\
9   & 33.3 & 1.0$\times$10$^6$ &2.44 & ~3.38 &~4.00  & ~4.17  & 2.00 &-\\
8   & 37.5 & 7.9$\times$10$^5$ &2.56 & ~3.57 &~4.44  & ~4.47  & 2.09 &-\\
7   & 42.9 & 6.0$\times$10$^5$ &2.72 & ~3.81 &~5.05  & ~4.86  & 2.20 &-\\
6   & 50.0 & 4.4$\times$10$^5$ &2.90 & ~4.10 &~6.00  & ~5.38  & 2.33 &-\\
5   & 60.0 & 3.1$\times$10$^5$ &3.13 & ~4.46 &~7.61  & ~6.06  & 2.49 &-\\
4   & 75.0 & 2.0$\times$10$^5$ &3.49 & ~4.96 &11.46  & ~7.09  & 2.72 & 1.02\\
3   & 100.0~ & 1.1$\times$10$^5$ &4.07 & ~5.75 &$>$\,30& ~8.88  & 3.07 & 1.37\\
2   & 150.0~ & 4.9$\times$10$^4$ &5.18 & ~7.07 &$>>$   & 12.17  & 3.69 & 2.06\\
1   & 300.0~ & 1.2$\times$10$^4$ &8.58 & 10.36 &$>>$   & 21.30  & 5.42 & 4.16\\
\hline
\end{tabular}
\mbox{{\bf a)} $n_0$=4.5 cm$^{-3}$ at 1 AU} \\
\end{table}

We have calculated atmospheric heights for some plasma frequencies 
using the basic (one-fold) model by Saito (equatorial densities), 
the hybrid model by Vr\u{s}nak, Magdaleni\u{c}, and Zlobec, 
a two-fold Newkirk model (approximation for active region corona), 
a ten-fold Saito model (approximation for streamer densities), 
the Leblanc, Dulk, and Bougeret model, and the IP model with an 
average electron density of 4.5 cm$^{-3}$ at \mbox{1 AU}, 
see Table \ref{table1}. 
The observed electron densities near \mbox{1 AU} before the shock 
arrival for our three analysed events are listed in Table \ref{table2}.  

\begin{table}[!hb]
\caption{Estimated electron density near 1 AU, around the times of the three events}
\label{table2}
\begin{tabular}{lcl}
\hline
\smallskip
Time & $n_0$       & Note  \\
     & (cm$^{-3}$) &       \\
\hline
27 Oct. 2003 22:00 UT & 0.9 & ACE (coronal hole prior to any ICMEs) \\
29 Oct. 2003 04:00 UT & 5.1 & ACE (right before first shock, CME launched \\
                      &     & on 28th)\\
                      & 4.0 & {\it Geotail} \\
30 Oct. 2003 16:00 UT & 1.0 & ACE (right before second shock, CME lauched\\
                      &     & on 29th)\\ 
\hline
06 Nov. 2004 10:00 UT & 4.5 & ACE (prior to first ICME)\\
09 Nov. 2004 06:00 UT & 1.3 & ACE (right before first shock of the second ICME)\\
\hline
16 Jan. 2005 00:00 UT & 4.5 & ACE (small coronal hole prior to first CME)\\
21 Jan. 2005 07:00 UT & 3.8 & ACE (right before shock)\\
\hline
\end{tabular}
\end{table}

To obtain CME velocity estimates from radio observations we make the 
basic assumption that the DH (IP) type II burst emission is formed by 
accelerated electrons near the bow shock at the leading edge (nose) 
of a CME. A futher assumption is that the outermost bright structures
in white light, projected on the plane-of-the-sky, represent the CME 
height with sufficient accuracy, although we can expect a certain offset 
between the burst driver and the bow shock \cite{russell02}. We use the 
observed (projected) CME heights as a constraint for the 
atmospheric-density models, and select the ones that fit best with 
the white-light CME observations. 
As an example for the usability of this method we refer to an event 
presented by Ciaravella {\it et al.} (\citeyear{ciaravella05}), 
where the CME leading edge was identified as the shock front with 
SOHO UVCS observations on 3 March 2000, at 02:19 UT. 
Simultaneous radio spectral observations from HiRAS show a type 
II burst. We calculated the radio source height using the fundamental 
emission at 40 MHz (the HiRAS dynamic spectrum is available at the 
HiRAS webpage). When the shock was observed by UVCS at a height of  
1.7 {\rsun}, the best-fit radio source height was 1.78 {\rsun}, 
calculated with the ``hybrid'' density model. 

In determining the burst driver speed a further complication is presented 
by direction of the motion against the density gradient. If the driver is 
propagating along the radial density gradient, the observed velocity 
is the true shock velocity, but a 45\degree angle ($\theta$) between 
the directions increases the true velocity by 1/$\cos \theta$, to 
about 1.4 times the observed velocity ({\it i.e.} the true distance is 
longer). Normally the direction of propagation does not differ much 
from the radially-decreasing density, so this can be taken as the
upper limit for velocity correction. 
   
At decimetric wavelengths, a common method for determining shock speeds 
is to use the observed frequency drift rate and derive the density scale 
height. If we differentiate the expressions for plasma frequency and 
electron density and use the relation $f \sim \sqrt{n}$, we get an
expression for the shock velocity ($v$) 
  \begin{equation}
   v = 2 \frac{1}{f} \frac{{\rm d}f}{{\rm d}t} H ,  \, \, \textrm{where} \, \,
   H = \left( \frac{1}{n} \frac{{\rm d}n}{{\rm d}r} \right)^{-1}.
  \end{equation}
The density scale height ($H$) and shock velocity ($v$) are expressed in km 
and km s$^{-1}$, observing frequency (fundamental emission, $f$) in
MHz, and the measured frequency drift d$f$/d$t$ (from a logarithmic
spectral plot where it should appear as a straight line) in MHz s$^{-1}$. 

For the local density scale height ($H_L$, in km) we can use a barometric 
isothermal density law for unmagnetized plasma, and the local electron density 
($n_e$) can be calculated from a reference density ($N_e$) at the base of the
corona:  
  \begin{equation}
   n_e = N_e \, \textrm{exp} \biggl(-\frac{696\,000}{H_L}\left
    (1 - \frac{1}{R}\right)\biggr), 
  \end{equation}
where $R$ is the estimated heliocentric distance (observed or derived from
a density model) in units of the solar radius {\rsun}. Details of this 
method can be found in, {\it e.g.}, D\'emoulin and Klein (\citeyear{dem00}).
In general, the method tends to give shock speeds smaller than those 
obtained directly from the density models described above. The barometric 
density law ceases to work at large heights, where the solar-wind densities 
become a dominant factor. This density change happens at frequencies  
near 5 MHz, and for this reason we use the scale-height method only for
the decimetric emission on 20 January 2005 in this paper.

\subsection{INTERPLANETARY SCINTILLATION OBSERVATIONS}

The propagation signatures of a CME in the space outside the
LASCO field of view can be obtained from the remote-sensing
Interplanetary Scintillation (IPS) technique ({\it e.g.}, Manoharan
{\it et al.} \citeyear{mano01}; Tokumaru {\it et al.} \citeyear{tokumaru05}).  
The IPS method exploits the scattering of radiation from distant 
radio sources (quasars, galaxies, {\it etc.}) by the density 
irregularities in the solar wind. The normalized scintillation 
index ({\it g}), where 
  \begin{equation}
   g = \frac{\rm observed\,\,\, scintillation\,\,\, index}
   {\rm expected\,\,\, average\,\,\, scintillation\,\,\, index}
   \end{equation}
can differentiate between the ambient background solar wind flow
and the excessive level of density turbulence associated with the
IP transients. The values of {\it g } close to unity represent the
undisturbed or background condition of the solar wind and values
{\it g} $>$ 1 and $<$ 1, respectively, indicate the increase and
decrease of density turbulence level in the IP medium. 

Since a propagating CME produces an excess of density turbulence by 
the compression of the solar wind ({\it i.e.}, the sheath region) between the 
shock and the driver ({\it i.e.}, the CME), the portion of the IP medium 
in front of the CME can be identified and tracked in the Sun-Earth 
distance with the help of scintillation images. Recent IPS studies have
revealed the increase of the linear size of the CME with distance,
suggesting the pressure balance is maintained between the CME cloud
and the ambient solar wind, in which the CME is immersed. The
radial profiles of some of the events indicate that the internal
energy of the CME supports the propagation, {\it i.e.}, the expansion
(Manoharan {\it et al.}, \citeyear{mano00}; \citeyear{mano01}). 
Further, IPS data have been useful to quantify the force of 
interaction experienced by CMEs in the IP medium
before their arrival at the near-Earth environment (Manoharan {\it et al.}, 
\citeyear{mano01}; Manoharan, \citeyear{mano06}). In other words, 
the CMEs moving with speeds less than the background solar-wind speed 
are accelerated (or aided by the ambient wind). On the other hand, 
faster CMEs are decelerated due to the drag force encountered by 
them in the IP medium \cite{mano06}. Therefore, to assess the 
radial evolution of a CME before its arrival at 1 AU, measurements 
at several heliocentric distances are essential.

In the present study, the multi-point IPS measurements of CMEs
under investigation in the Sun-Earth space have been obtained
with the Ooty Radio Telescope, which is operated by the Radio
Astronomy Centre, Tata Institute of Fundamental Research, India.
The description of Ooty IPS observations and the method of data
reduction procedure have been given by Manoharan (\citeyear{mano06}) 
and references therein. At Ooty, regular monitoring of scintillation 
is made each day for about 700 to 900 radio sources.
The CME events on 7\,--\,9 November 2004 and 20\,--\,21 January 2005
have been covered by the Ooty IPS measurements. However, due to the
annual maintenance of the Ooty Radio Telescope, the event on
28\,--\,29 October 2003 has not been observed by the IPS method. 
It should be noted that the IPS measurements are sensitive
to solar-wind structures of a CME propagating perpendicular
to the line of sight to the radio source. An IPS image,
therefore constructed using the data obtained from a large
number of lines of sight passing through different parts of
the CME, will provide the solar offset in the sky plane
similar to the measurement obtained from a LASCO image.

\section{Analysis and Results from the Three Halo CME Events}

\subsection{28 OCTOBER 2003}

\begin{figure}
\begin{center}
\includegraphics[width=0.3\textwidth]{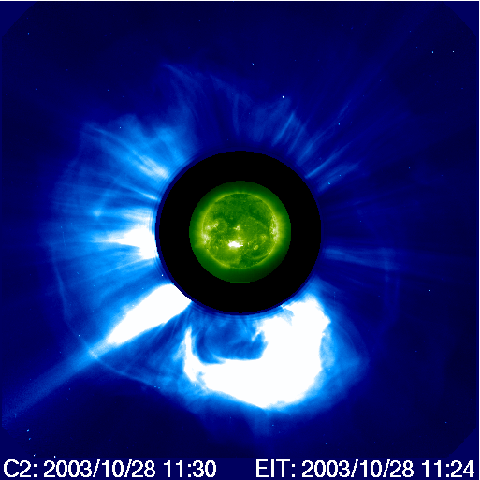}
\includegraphics[width=0.3\textwidth]{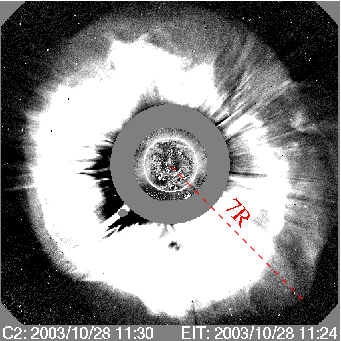}
\includegraphics[width=0.3\textwidth]{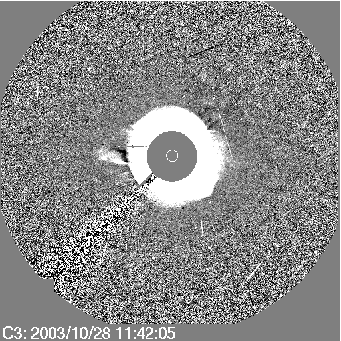}
\caption{
{On the left}: 
LASCO C2 image of the halo CME at 11:30 UT on 28 October 2003, 
with insert EIT disk image at 11:24 UT. 
{In the middle:} Corresponding running-difference images.  
The outermost CME front is located at a heliocentric distance 
of 7 {\rsun} at 11:30 UT. 
{On the right}: LASCO C3 running-difference image at 11:42 UT,
showing a symmetric halo structure.
} 
\label{fig2a}      
\end{center}
\end{figure}

A full halo-type CME was observed at 11:30 UT in LASCO C2 (leading front 
at distance 7 {\rsun} towards the southwest) and at 11:42 UT in C3 
(leading front at distance 10 {\rsun} in the same direction), see
Figure~\ref{fig2a}. These two measurements give a CME speed of 2900 
km s$^{-1}$. In the LASCO CME Catalog the CME fronts are measured at 
a different angle towards the North where the heights are lower: 
5.84 {\rsun} at 11:30 UT and 8.77 {\rsun} at 11:42 UT. The CME speeds 
from the catalogue are 2460 km s$^{-1}$ (linear fit to all data points) 
and 2700 km s$^{-1}$ (second-order fit, speed near 11:30 UT). The 
second-order fit indicates that the CME is decelerating, and the 
speed is estimated to drop to 2000 km s$^{-1}$ near 30 {\rsun}. 

A GOES X17-class flare started around 11:00 UT, superposed on an 
earlier event. Although this event is a classic halo CME, only three 
useable LASCO frames are available. For these the velocity asymmetry 
is not measurable and so the model of Micha\l ek, Gopalswamy, and
Yashiro (\citeyear{michalek03}) can not be applied for the estimation 
of radial velocity.

\begin{figure}
\begin{center}
\includegraphics[width=0.9\textwidth]{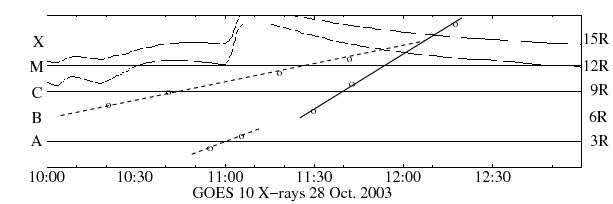}
\caption{GOES X-ray flux curves in the 1.5\,--\,12.5 keV and 
3\,--\,25 keV energy bands, and the estimated heights of the 
halo CME on 28 October 2003. 
The height--time trajectory is marked with a solid line. An
impulsive rise in soft X-rays was observed near 11:00 UT, although
gradual rise had been recorded from 09:50 UT onwards. The flare was
classified as GOES X17 class. Two other CMEs preceded the halo CME;
their heights and times are marked with dashed lines.  
(Data from the LASCO CME Catalog.)}
\label{fig2b}
\end{center}      
\end{figure}

A bright loop front directed towards the southeast was observed before 
the halo CME, at 10:54 and 11:06 UT. The projected speed of this loop 
front, classified as a partial halo CME in the LASCO catalogue, 
was 1054 km s$^{-1}$, which is less than half of the speed of the main 
halo event. Due to the difference in speed and direction, it is not clear 
if these two ejections are related. However, the halo event observed 
at 11:30 UT almost certainly interacted with the material from the 
preceding event; see the height--time diagram in Figure~\ref{fig2b}.

\begin{figure}
\begin{center}
\includegraphics[width=0.9\textwidth]{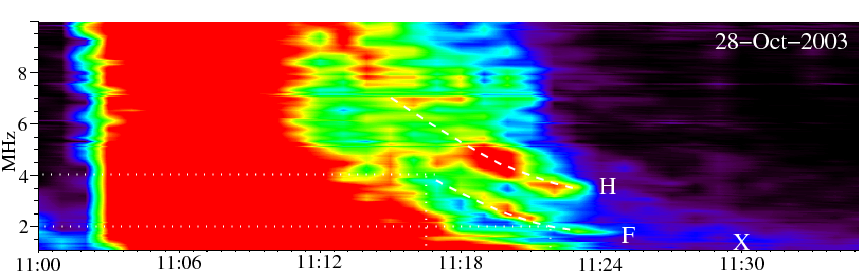}
\caption{{\it Wind} WAVES RAD2 data at 1\,--\,10 MHz from 
28 October 2003.
Fundamental (``F'') and harmonic (``H'') emission lanes of the DH 
type II burst are indicated in the plot. The ``X'' at 11:30 UT marks 
the observed distance of the CME front from the Sun centre, 
7 {\rsun}, that corresponds to the frequency of 1.3 MHz using the 
Saito (1970) density model. The times and frequencies used for the 
speed calculation are indicated with dotted lines. }
\label{fig2c}      
\end{center}
\end{figure}

\begin{table}[!hb]
\caption{Estimated radio burst velocities on 28 October, 2003}
\label{table3}
\begin{tabular}{lcccccc}
\hline
UT & $f_p$ & $n_e$ & $h$ Saito & $h$ Hybrid & $h$ Leblanc$^{\bf a}$ & $h$ IP$^{\bf a}$\\
   &       & (cm$^{-3}$) &(\rsun) &(\rsun) & (\rsun) & (\rsun)\\
\hline
11:16:30 & 4~ MHz & 2.0$\times$10$^5$ & 3.49 & 4.96 &2.77 & 1.07 \\
11:21:54 & 2~ MHz & 4.9$\times$10$^4$ & 5.18 & 7.07 &3.78 & 2.17 \\
Velocity &       &    & 3630 km/s & 4530 km/s & 2195 km/s & 2365 km/s \\
11:30 & ~1.3 MHz$^{\bf b}$ & 2.1$\times$10$^4$  & 7.0~  & & & \\
Velocity &       &    & 2580 km/s & & & \\
\hline
\end{tabular} \\
\mbox{{\bf a)} $n_0$=5.0 cm$^{-3}$ at 1 AU} \\
{\bf b)} Frequency corresponding to the observed LASCO CME height\\
{\bf Constraints}: Observation of CME front at 11:30 UT: 7.0 \rsun 
(this study), 5.84 \rsun (CME Catalog) \\
{\bf LASCO velocities}:\\
CME front (plane-of-the-sky) 11:30-11:42 UT: 2900 km/s \\
CME front (CME Catalog, second-order fit) 11:30 UT: 2700 km/s 
\end{table}

At decimetric-metric wavelengths no clear type II emission is observed
(see, {\it e.g.}, dynamic spectra from IZMIRAN, at their webpage). At DH
wavelengths, {\it Wind} WAVES RAD2 observed a type II burst with fundamental 
and second-harmonic emission. The burst appears in the spectrum at 11:11 UT 
near 14 MHz, which is also the upper limit for the observing
frequency. Emission then drifts to the lower frequencies, at a rate 
of $\sim$0.01 MHz s$^{-1}$.  
At 11:16:30 UT the fundamental emission lane is at 4 MHz, see 
Figure~\ref{fig2c}, which corresponds to a plasma density of 2.0$\times$10$^5$ 
cm$^{-3}$. The density model by Saito gives a corresponding atmospheric 
height of 3.49 {\rsun}, which fits quite well with the estimated CME height
of 3.5 {\rsun} at that time (backwards extrapolation from the observed 
CME leading edge heights at 11:30 and 11:42 UT). For comparison, 
the hybrid, two-fold Newkirk, and ten-fold Saito models give unrealistically 
large heights, and the radio source heights given by the Leblanc and IP 
models are 0.8\,--\,2.5 {\rsun} lower than the estimated CME front height
on the plane of the sky. 
The ``X'' in Figure~\ref{fig2c} at 11:30 UT marks the observed 
height of the LASCO CME front (7 {\rsun}), corresponding to the 
frequency of 1.3 MHz in the Saito density model.
 
The derived shock speeds from the radio observations using the Saito 
density model, which gives best correspondence with height, are 3630 
km s$^{-1}$ between 11:16:30 and 11:21:54 UT, and 2580 km s$^{-1}$ 
between 11:21:54 and 11:30 UT, see Table \ref{table3}. The type-II-burst 
lane is not fully visible at 11:30 UT, but as Figure~\ref{fig2c} shows,
the frequency of 1.3 MHz corresponds well to the estimated continuation
of the type II lane.  
The CME speeds from the LASCO observations, approximately 2900 km s$^{-1}$
between the observations at 11:30 and 11:42 UT, and 2700 km s$^{-1}$
from the second-order fit in the LASCO CME Catalog, are not too far from 
the radio type II speeds calculated with the Saito density model. 
It seems evident that both the white-light CME and the shock driving 
the type II emission are decelerating. We note that a relatively faint 
streamer structure was observed towards the Southwest, which could 
affect density estimates. No interplanetary scintillation 
observations were available for this event.

\subsection{7 NOVEMBER 2004}

The halo-CME front was first observed at a height of 5.7 {\rsun} in 
the LASCO C2 image at 16:54 UT on 7 November 2004. The front was next 
visible in C3 at 17:18:05 UT at a height 9.6 {\rsun} (LASCO CME Catalog
and our analysis).
The CME propagated towards the North, with a projected plane-of-the-sky 
speed of 1890 km s$^{-1}$ (calculated from the first two observations). 
The CME speeds from the catalogue are 1760 km s$^{-1}$ (linear fit to 
all data points) and 1850 km s$^{-1}$ (second-order fit, speed near 
16:54 UT). The second-order fit indicates that the CME was decelerating  
at 20 m s$^{-2}$, and the speed is estimated to drop to 1600 km s$^{-1}$ 
near 30 {\rsun}. 

The CME was preceded by a X2.0-class flare that started at 15:42 UT, 
more than an hour before the CME was first observed. However, the first 
halo CME observation at 16:54 UT was preceded by two bad frames of LASCO 
C2 data, at 16:06 and 16:30 UT. 
The X2.0 flare was a complex event involving at least two flares in 
the active region AR10696, and the eruption of a trans-equatorial 
filament. The CME appeared to be launched by the flare associated with 
the last GOES flare peak between 16:20 and 16:54 UT (Harra {\it et al.}, 
\citeyear{harra07}, in this issue).

\begin{figure}[!h]
\begin{center}
\includegraphics[width=0.3\textwidth]{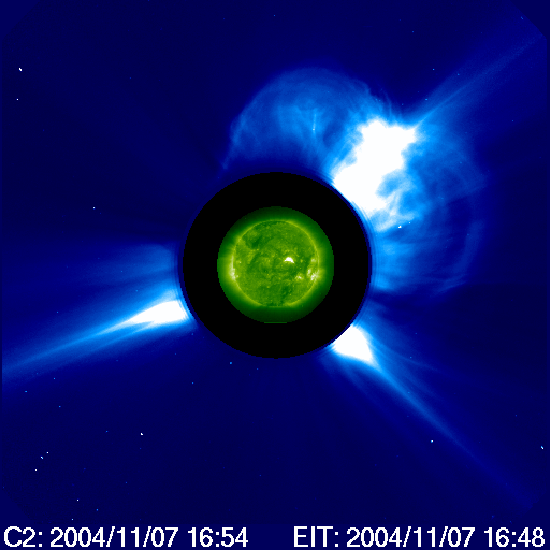}
\includegraphics[width=0.3\textwidth]{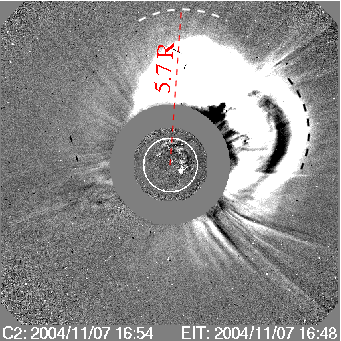}
\includegraphics[width=0.3\textwidth]{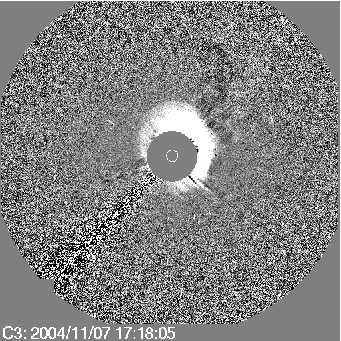}
\caption{
{On the left:}
LASCO C2 observation of the halo CME at 16:54 UT on 7 November 2004, 
with EIT insert image at 16:48 UT. 
{In the middle:} 
Corresponding running-difference images. The outermost CME 
front is located northward, near 5.7 \rsun (white dashed line). 
The black dashed line on the West side of the Sun marks the location 
of the earlier erupted, slower CME front near 4.9 \rsun. 
{On the right:} 
LASCO C3 running-difference image at 17:18 UT shows the asymmetric 
``halo'' structure.
}
\label{fig3a}      
\end{center}
\end{figure}

\begin{figure}[!h]
\begin{center}
\includegraphics[width=0.9\textwidth]{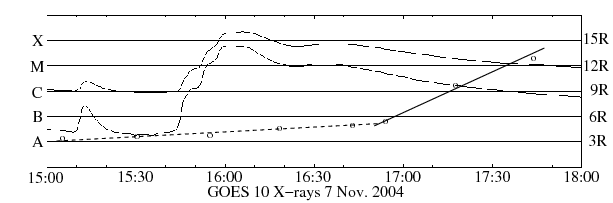}
\caption{GOES X-ray flux curves in the 1.5\,--\,12.5 keV and 3\,--\,25
keV energy bands, and the estimated heights of the halo CME on 
7 November 2004. The height--time trajectory is marked with a solid
line. A GOES X2.0-class flare was observed to start at 15:42 UT, and 
the halo CME was first observed at 16:54 UT.  
A preceding CME, listed by the LASCO CME Catalog, is also shown (heights
and times are indicated with a dashed line). LASCO C2 had bad frames 
at 16:06 and 16:30 UT, and the heights given here for the first, slower 
CME at 16:18 and 16:42 UT are from LASCO C3, which blocks the lower 
altitudes.
} 
\label{fig3b}      
\end{center}
\end{figure}

A slow CME (225 km s$^{-1}$) was observed to propagate towards the West 
during the X-class flare, and in the C2 image at 16:54 UT in 
Figure~\ref{fig3a} both CME fronts can be seen simultaneously. The 
slow CME has a height of 4.9 {\rsun} at that time. The later LASCO C3 
running-difference image in Figure~\ref{fig3a} at 17:18 UT shows the 
full halo. Figure \ref{fig3b} shows the height--time diagram for these 
structures, with the GOES time profile. 

With the availability of two LASCO C2 and six LASCO C3 frames that
are useable, the  Micha\l ek, Gopalswamy, and Yashiro
(\citeyear{michalek03}) cone model (see Section 2.1) was applied for 
this event. Given the identification of the launch site (active region 
that produced the X-class flare, at N09\,W17), the value of 
$r$/{\rsun} = 0.287 yields $\gamma$ = 73.3\degree. 
The values of the plane-of-the-sky velocities at the appropriate opposite 
limbs are 1146 and 660 km s$^{-1}$, respectively. These, in turn, lead 
to values of $\alpha$ = 78.1\degree and $V$ = 1488 km s$^{-1}$ using 
Equations (3) and (4) from Micha\l ek, Gopalswamy, and Yashiro.

The $\gamma$ and $\alpha$ angles determine the propagation direction
and angular width of the CME, and we can use them to deproject the true
height. In the coronagraph images we see the projection of the CME 
cone's outer edge, which has an angle of $\gamma - \alpha/2$ with the 
plane of the sky. To obtain the true heliocentric height we 
have to substract $r$/{\rsun} from the projected height divided by 
$\cos(\gamma-\alpha/2)$. For the 7 November 2004 CME this method 
yields 1.2\,$\times$\,the projected height\,$-$\,0.287 {\rsun}.
If the projected height of the CME front at 16:54 UT was 5.7 {\rsun},
the true height would have been about 6.5 {\rsun}.
Figure~\ref{fig3a} shows that we have measured the projected height
from the slightly brightened edge towards the North. If we had made 
the measurement from the outermost bright bulk the height would be 
considerably less.
 
\begin{figure}
\begin{center}
\includegraphics[width=0.9\textwidth]{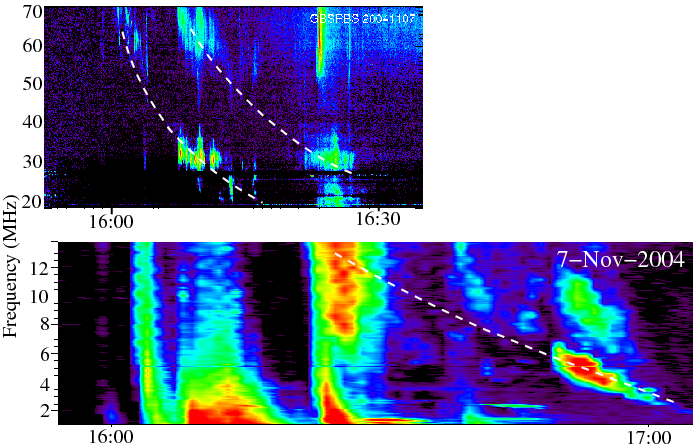}
\caption{GBSRBS dynamic spectrum at 20\,--\,70 MHz (top) 
and {\it Wind} WAVES spectrum at 1.1\,--\,14 MHz (bottom) 
on 7 November 2004. The metric type II burst emission observed by 
GBSRBS can be interpreted as having a continuation in the DH range.} 
\label{fig3c}
\end{center}      
\end{figure}

The dynamic spectrum at metric wavelengths (20\,--\,70 MHz) in 
Figure \ref{fig3c} from GBSRBS reveals several bursts that can be 
connected with the bursts in the {\it Wind} WAVES spectrum at DH wavelengths. 
However, the complex event with several CME structures makes the 
identification of type II lanes difficult. Along the shock trajectory
atmospheric densities can vary, which can explain the fast frequency drifts 
and patchy type II lanes. If the metric and DH burst emissions in this case   
are related, this could be one of the few observations of a coronal 
shock wave propagating into the IP medium.

\begin{figure}
\begin{center}
\includegraphics[width=0.9\textwidth]{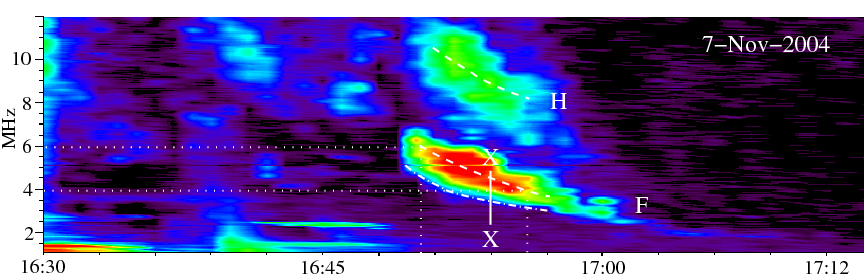}
\caption{{\it Wind} WAVES RAD2 data at 1.1\,--\,12 MHz from 7 November 2004.
The two lanes can be interpreted as fundamental and second-harmonic 
emission of the DH type II burst (indicated with dashed lines in the plot). 
The observed plane-of-the-sky distance of the CME front at 16:54 UT 
was 5.7 \rsun, which corresponds to a frequency of 1.7 MHz using the 
Saito (1970) density model and 5.5 MHz using the ten-fold Saito model. 
This frequency range is marked with ``X''s in the plot. We made height 
calculations using both the center (at 4.5 MHz at 16:54 UT) and the 
leading edge (at 3.5 MHz) of the emission lane at the fundamental 
(leading edge marked with a dash-dotted line).}
\label{fig3d}      
\end{center}
\end{figure}

As the bursts in the {\it Wind} WAVES spectrum are patchy and consist of
separate ``blobs'', we have selected one continuous lane  of the clearly 
fundamental emission for analysis, shown in Figure~\ref{fig3d}.
The LASCO CME observation at 16:54 UT is during this part of the 
type II burst, and we can compare the source heights directly without having 
to extrapolate for the CME front location. The center of the type II
fundamental lane at 16:54 UT is at 4.5 MHz and the low frequency edge
at 3.5 MHz. The observed CME height of 5.7 {\rsun} at that time corresponds 
to a frequency of 1.7 MHz using the Saito density model (equatorial region), 
and 5.5 MHz using the ten-fold Saito model (high density loops or streamers). 
This means that the ``true'' density can be in between these two model 
densities, and nearer the high-density Saito model. An in-between model, 
a seven-fold Saito, would produce radio emission at 4.5 MHz near height 
5.7 {\rsun}. 

The derived burst-driver speed using the seven-fold Saito density model
is 2955 km s$^{-1}$. This figure is high compared to the white-light 
observations of the CME front velocity. Other density models give 
lower burst speeds, from 800 km s$^{-1}$ (Leblanc model) to 1765 km s$^{-1}$ 
(hybrid model), see Table \ref{table4}. Using the leading edge of the 
fundamental emission lane gives slightly larger burst speeds, 
but the deprojected CME height agrees with the height given by 
the seven-fold Saito density model. 

The CME velocity derived from the cone model (1490 km s$^{-1}$) is less
than the plane-of-the-sky speed, but Micha\l ek, Gopalswamy, and 
Yashiro do report smaller radial speeds in some cases. This speed 
agrees with the speed based on the Saito density model but the height 
does not.  The best match with the height and speed comes from using 
the hybrid model. 

\begin{figure}[!ht]
\begin{center}
\includegraphics[width=0.4\textwidth]{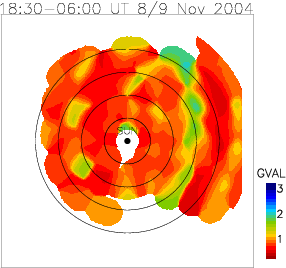}
\includegraphics[width=0.5\textwidth]{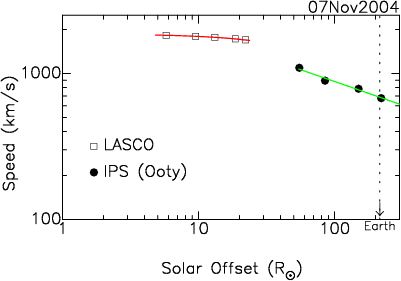}
\caption{Ooty scintillation image (left) and CME speeds (right)
from the SOHO LASCO CME Catalog and the measurements from the Ooty 
interplanetary scintillation observations during 7\,--\,9 November 2004.} 
\label{fig3e}      
\end{center}
\end{figure}
 
Figure~\ref{fig3e} shows an IPS image of the IP medium between 18:30 UT 
(8 November) and 06:00 UT (9 November). In this ``PA--heliocentric distance'' 
image, the North is at the top and East is to the left. The concentric 
circles are of radii, 50, 100, 150, 250 {\rsun}. 
The red color code indicates the background (ambient) solar wind. 
The observing time increases from the West of the image (right side 
of the plot) to the East (left). 

\begin{table}[h!]
\caption{Estimated radio burst velocities on 7 November, 2004}
\label{table4}
\begin{tabular}{ccccccc}
\hline
UT&$f_p$&$n_e$ & $h$ Hybrid  & $h$ 7-fold Saito & $h$ Saito & $h$ Leblanc$^{\bf a}$ \\
  &     & (cm$^{-3}$) & (\rsun) &(\rsun) & (\rsun) & (\rsun) \\
\hline
16:50:19 & 6~ MHz & 4.4$\times$10$^5$ & 4.10 & 4.79 & 2.90  & 2.33 \\
16:55:58 & 4~ MHz & 1.1$\times$10$^5$ & 4.96 & 6.23 &  3.49  & 2.72 \\
\multicolumn{3}{l}{Velocity (lane center)} & 1765 km/s &  2955 km/s & 1210 km/s & 800 km/s\\
16:50:19 & 4.5 MHz & 2.5$\times$10$^5$ & 4.70 & 5.76 &  3.30  & 2.59 \\
16:54:00 & 3.5 MHz & 1.5$\times$10$^5$ & 5.31 & 6.88 &  3.73  & 2.87 \\
\multicolumn{3}{l}{Velocity (lane edge)} & 1920 km/s & 3530 km/s & 1355 km/s & 900 km/s\\
\hline
\end{tabular} \\
{\bf a)} $n_0$=4.5 cm$^{-3}$ at 1 AU \\
{\bf Constraints}: observation of CME front at 16:54 UT at 5.7 \rsun, deprojected
height 6.5 \rsun \\
{\bf LASCO and IPS velocities}:\\
CME front (cone model): 1490 km/s \\
CME front (plane-of-the-sky) 16:54--17:18 UT: 1890 km/s\\
CME front (CME Catalog, second-order fit) 18:40 UT: 1700 km/s \\
CME front (IPS extrapolation) 18:40 UT: 1460 km/s \\
\end{table}

The region of enhanced density turbulence associated
with the CME is seen at $\sim$150 {\rsun}.  The adjacent plot
shows the plane-of-the-sky speeds of the CME in the LASCO and IPS 
fields of view. It is interesting to note that the speed profile 
of the CME in the IPS field of view, when extrapolated to the 
LASCO field of view, indicates a speed of $\sim$1460 km s$^{-1}$ 
at the heliocentric height of $\sim$22 {\rsun}, which is about 
the same as the speed deduced from the cone model, but about 
200 km s$^{-1}$ less than the LASCO CME Catalog plane-of-the-sky 
speed. It is evident from the extended region of enhanced density 
turbulence seen in the IPS image that the CME had gone through 
structural changes with increasing heliocentric distance, that
decreased the speed.

\subsection{20 JANUARY 2005}

\begin{figure}
\begin{center}
\includegraphics[width=0.3\textwidth]{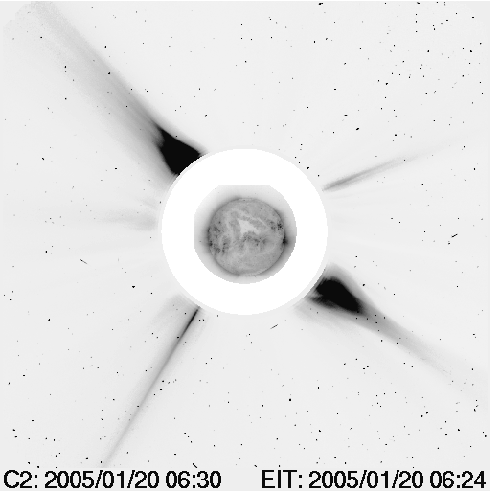}
\includegraphics[width=0.3\textwidth]{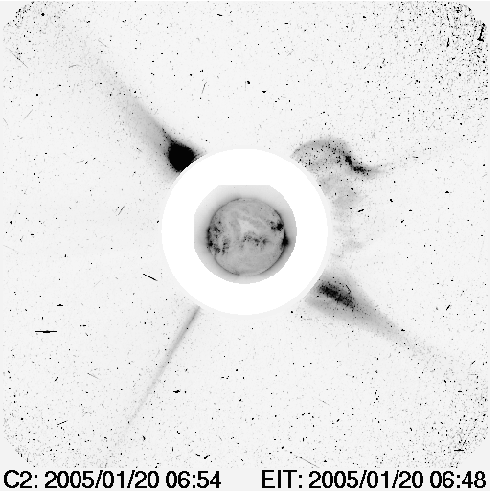}
\includegraphics[width=0.3\textwidth]{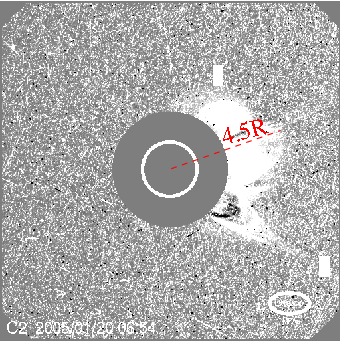}
\caption{LASCO C2 observations of the halo CME on 20 January 2005. 
At 06:30 UT (on the left) the CME is not yet visible. At 06:54 UT 
(in the middle) the CME is observed in the nortwest. The insert 
EIT images show the solar disk at 06:24 and 06:48 UT, respectively. 
The outermost CME front is located near 4.5 {\rsun} at 06:54 UT, 
measured from the centre of the Sun (on the right). The left 
and middle C2 images have reversed color scales, and the image on 
the right is a C2 running-difference image. All images suffer from
intense particle hits.  
}
\label{fig4a}      
\end{center}
\end{figure}

\begin{figure}[!h]
\begin{center}
\includegraphics[width=0.9\textwidth]{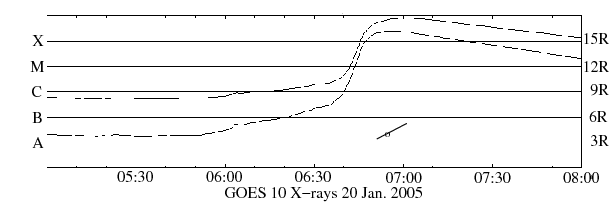}
\caption{GOES X-ray flux curves in the 1.5\,--\,12.5 keV and 
3\,--\,25 keV energy bands, with the only reliable height--time 
data point of the halo CME on 20 January 2005. An impulsive rise 
in soft X-rays was observed near 06:40 UT, with a brightening 
in H$\alpha$. The flare was classified as GOES class X7.1. } 
\label{fig4b}      
\end{center}
\end{figure}

The LASCO images for the 20 January 2005 event were limited to
only one undamaged image taken at 06:54 UT, see Figure~\ref{fig4a}.
Hence the cone model cannot be used for this event. The CME loop was
directed to the Northwest, and a streamer structure is visible in
the Southwest. The CME front was at a height of \mbox{$\sim$4.5 {\rsun}} 
at 06:54 UT. A GOES X7.1-class flare showed an impulsive rise near 
06:40 UT (Figure~\ref{fig4b}), which caused intense particle hits 
and saturation effects in all instruments. Thus the speed of 
the white-light CME cannot be estimated reliably from the LASCO 
observations.

\begin{figure}
\begin{center}
\includegraphics[width=0.9\textwidth]{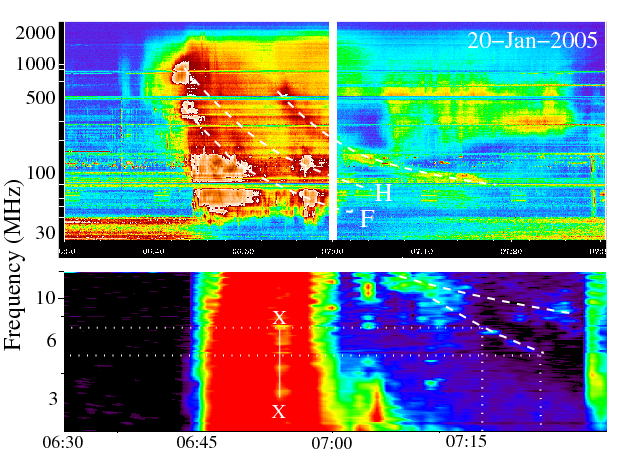}
\caption{HiRAS decimetric-metric spectrum (top) shows several 
metric type II burst lanes associated with the 20 January 2005 event. 
The {\it Wind} WAVES RAD2 dynamic spectrum (bottom) shows at least 
two possible type-II-burst lanes at decametric wavelengths. 
The type-II-burst lanes are indicated with white dashed lines in the spectral
plots. The ``X''s at 06:54 UT mark the observed height of the
white-light CME front at 4.5 {\rsun} which corresponds to the frequency of 
2.5 MHz using the Saito (1970) density model and to 7.9 MHz using the 
ten-fold Saito model. We estimated the burst speed for one of the 
type II lanes (the times and frequencies used for the
calculations are indicated with dotted lines).}
\label{fig4c}
\end{center}      
\end{figure}

In the decimetric-metric dynamic spectrum from HiRAS, several type 
II burst lanes can be observed. At 06:43 UT a patchy 
fundamental--second-harmonic emission lane pair is observed to 
start near 800 and 400 MHz, see spectrum at the top of Figure~\ref{fig4c}. 
At 06:54 UT a single type II lane is observed to start near 600 MHz. 
The heights of these radio sources are well below the height of the 
white-light CME front observed at 06:54 UT.  
Estimates for the burst heights can only be calculated with 
high-density models like the ten-fold Saito, since other density models 
do not have a solution for the high densities/high starting frequencies   
(Table \ref{table1}). The derived speeds for these decimetric bursts, 
using the ten-fold Saito densities, are around 650 km s$^{-1}$. 
The density scale height method can also be used in this case.
If we assume a base density of 10$^9$ cm$^{-3}$ and a heliocentric 
distance of 1.3 {\rsun} for an emission structure at 100 MHz 
(in agreement with the hybrid and Newkirk density models), we get a 
local density scale height of 75\,800 km. The frequency drift of the 
type II burst is around 0.6 MHz s$^{-1}$ between 400 and 100 MHz (the
usual metric type II burst drifts are between 0.1 and 1.0 MHz s$^{-1}$, 
see Nelson and Melrose, \citeyear{nelson85}) and the obtained burst speed 
is around 940 km s$^{-1}$. 

The DH type II bursts observed with the {\it Wind} WAVES instrument are faint 
and patchy. The two type-II-burst lanes, shown in the dynamic spectrum 
at the bottom of Figure~\ref{fig4c} and indicated with dashed lines, 
could both be a continuation of the metric emission, but due to the 
observational gap in the frequency range these are difficult to interpret. 
We calculated the speed for one of the burst lanes, the selected times 
and frequencies are indicated with dotted lines, and the speed varies 
from 750 (Saito model) to 4690 km s$^{-1}$ (two-fold Newkirk model), 
see Table \ref{table5}. These highly different numbers reflect the need to 
have a reference height for the propagating shock.  

The observed LASCO CME front near 4.5 {\rsun} at 06:54 UT cannot be used 
as a constraint here as the only LASCO observation is well before 
the DH type II burst becomes visible, and the CME height corresponds to 
a much lower frequency than where the type II appears.
The frequency range corresponding to the CME height is marked with 
``X''s in the dynamic spectrum in Figure~\ref{fig4c}. This falls inside
intense continuum emission, that would hide any type II burst.  
We cannot associate the LASCO CME front with any of the metric type II 
bursts either, since this would require extremely high densities at 
large atmospheric heights, which are not very likely.

\begin{figure}
\begin{center}
\includegraphics[width=0.9\textwidth]{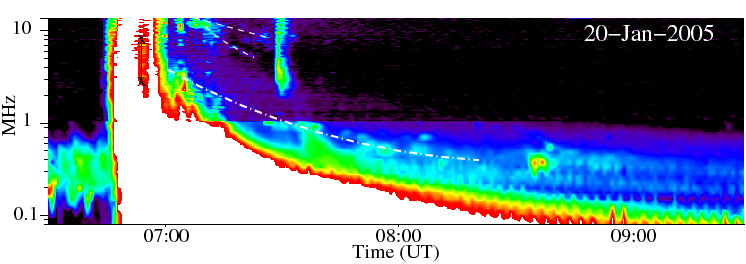}
\caption{{\it Wind} WAVES RAD1 and RAD2 dynamic spectra at 
06:30\,--\,09:30 UT on 20 January 2005 (80 kHz\,--\,14 MHz frequency 
range shown here) suggest that the CME-driven shock propagated already 
at hectometric wavelengths near 7 UT, see indicated burst lane 
(dash-dotted line). A clear continuous burst lane is not visible in 
the spectrum, but the individual short bursts appear at regular time 
intervals. The heights and times of this lane also agree with the 
white-light CME observations.}
\label{fig4d}
\end{center}      
\end{figure}

It is probable that if there was a DH type II burst associated with
the leading edge of the CME, it was hidden by the intense continuum
emission. Some indication of this is visible in the {\it Wind} WAVES RAD1
spectrum. A composite of the RAD1 and RAD2 observations at 
80 kHz\,--\,14 MHz is shown in Figure~\ref{fig4d}, which shows burst 
patches at hectometric wavelengths between 07 and 09 UT. The times and heights
along the indicated burst lane agree with the height of the white 
light CME at 06:54 UT (4.5 {\rsun}) if the hybrid density model is used, 
as emission near 6 MHz comes roughly from a height of 4.1 {\rsun}.
The burst speed, calculated with the hybrid density model from the emission 
heights along the patchy lane at 3.5 MHz (07 UT) and 500 kHz (08 UT) is 
1990 km s$^{-1}$, see Table \ref{table5}.    

\begin{figure}
\begin{center}
\includegraphics[width=0.4\textwidth]{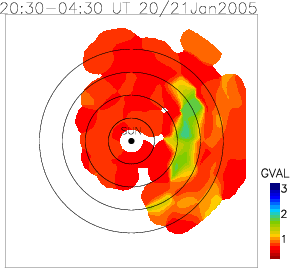}
\includegraphics[width=0.5\textwidth]{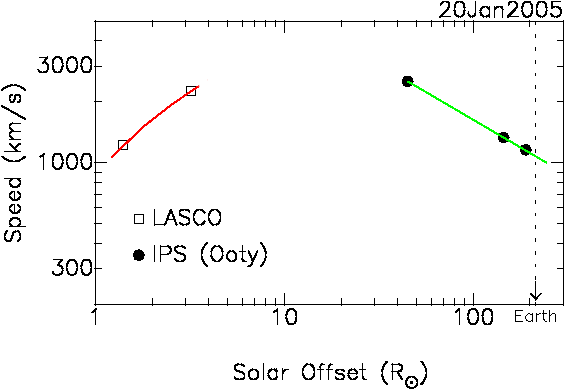}
\caption{Ooty scintillation image (left) and CME speeds (right)
from the estimations by Gopalswamy {\it et al.} (2005) and the 
measurements from the Ooty interplanetary scintillation observations 
during 20\,--\,21 January 2005.} 
\label{fig4e}      
\end{center}
\end{figure}

Since the LASCO images were contaminated by the intense particle event, 
Gopalswamy {\it et al.} (\citeyear{gopal05}) combined the usable LASCO C2 image
with signatures of the CME observed in earlier SOHO EIT data,
to derive a height--time plot. Comparison of structures visible in 
the EIT difference image at 06:36 UT and in the LASCO C2 difference 
image at 06:54 UT suggest a high speed, $\sim$ 2100 km s$^{-1}$. 
In Figure~\ref{fig4e} these data points are indicated by square symbols.
A second-order fit to these heights and times would give very high 
speeds later on.  

Although the CME was apparently moving fast, in the IPS field of 
view three measurements could be used to track the CME at heliocentric
distances $>$50 {\rsun}. In Figure~\ref{fig4e}, a representative Ooty IPS 
image of the CME is shown. The estimated speed of the CME in the IPS 
field of view, on the right in Figure~\ref{fig4e}, is consistent with 
the high speed derived by Gopalswamy {\it et al.} (\citeyear{gopal05}). 
The IPS measurements show that the CME decelerated from about 
2500 km s$^{-1}$ at a solar offset of 50 {\rsun} to $\sim$1000 km s$^{-1}$ 
at $\sim$1 AU. However the plot in Figure~\ref{fig4e} suggests that
the speed may have reached about 3000 km s$^{-1}$ in the interval 
3 {\rsun} to 50 {\rsun}. 

\begin{table}[!h]
\caption{Estimated radio burst velocities on 20 January, 2005}
\label{table5}
\begin{tabular}{ccccccc}
\hline
UT&$f_p$& $n_e$ & $h$ Saito & $h$ 10-fold & $h$ Hybrid & $h$ 2-fold \\
  &     &       &           & Saito       &            & Newkirk \\
  &     & (cm$^{-3}$) & (\rsun) & (\rsun) & (\rsun)  & (\rsun) \\
\hline
07:16:55 & 7~ MHz & 6.0$\times$10$^5$ & ~2.72 & 4.86 & ~3.81 & 5.05 \\
07:23:15 & 5~ MHz & 3.1$\times$10$^5$ & ~3.13 & 6.06 & ~4.46 & 7.61 \\
Velocity & & & 750 km/s & 2200 km/s & 1190 km/s & 4690 km/s \\
07:00~~  & ~3.5 MHz & 1.5$\times$10$^5$ & ~3.73 &  & ~5.31 &  \\ 
08:00~~  & 500 kHz~ & 3.1$\times$10$^3$ & 14.62 &  & 15.57 &  \\
Velocity & & &                       2105 km/s &  & 1990 km/s &  \\
\hline
\end{tabular}
{\bf Constraint}: observation of CME front at 06:54 UT at 4.5 \rsun \\
{\bf LASCO and IPS velocities}:\\
CME front (EIT and LASCO image) estimated: $>$2100 km/s \\
CME front (IPS) at 50 {\rsun}: 2500 km/s \\
\end{table}

\section{Discussion on the Three Halo CME Events}

The halo CMEs on 28 October 2003 and 7 November 2004 were both preceded 
by a slower CME. Both halo CMEs were found to be decelerating after 
initiation. During the 20 January 2005 event no evident CME
``interaction'' was observed, and there is some evidence to suggest that 
this CME was accelerating at low atmospheric heights.  
In all three events deceleration from the initial CME speed is 
evident from the arrival times of the magnetic clouds and the measured 
speeds near Earth, see Table \ref{table6}. The transit speeds of the 
7 November 2004 CME/magnetic cloud are discussed in more detail in 
Culhane {\it et al.} (\citeyear{culhane06}).
  
As all three CMEs were faster than the solar wind, the drag force exerted 
by the wind slowed them down considerably. Radio scintillation observations 
at heliocentric distances $>$50 {\rsun} on 7\,--\,9 November 2004 and 
20\,--\,21 January 2005 also record CME deceleration, at 3 and 
30 \mbox{m s$^{-2}$} respectively, but where this deceleration begins
is not so clear. 
On 7 November 2004 the white-light CME was observed to decelerate 
at 20 m s$^{-2}$, which is consistent with the expectation that the 
CME should decelerate more rapidly near the Sun where the density 
is high, since the drag force depends on the density and velocity.
The observed speeds near Earth from IPS measurements are in agreement 
with the measured speeds of the IP shocks and magnetic clouds. 

Wang {\it et al.} (\citeyear{wang05}) have made simulations of CME 
interactions, in a case where a slow CME is overtaken by a faster one, 
with the result that the faster CME is slowed down significantly by 
the blocking action of the preceding magnetic cloud. 
This implies that the travel time is dominated
by the slower cloud. Therefore it is possible that interaction with 
earlier, slower CMEs affected the speeds in the first two events. 
The simulations made by Lugaz, Manchester, and Gombosi
(\citeyear{lugaz05}) indicate that a CME can accelerate when it first 
meets the rear edge of a preceding CME (due to the density drop), 
and only decelerates later when it passes through the center and 
front of the earlier launched CME. 
Radio observations made at the time of CME acceleration 
would then give too high an instantaneous speed value. This could possibly
explain the higher radio-source (shock) velocity compared to the CME 
velocity on the 7 November 2004 event, where the CME and radio source 
heights are in good agreement.

\begin{table}[]
\caption{Estimated speeds for the magnetic clouds (MCs) and the IPS
 velocity \mbox{estimates near Earth} }
\label{table6}
\begin{tabular}{llcccl}
\hline
Launch         & Arrival  & Average & MC speed & Sun       & Notes \\
time           & time     & transit & near     & to        & \\
               &          & speed   & Earth    & Earth     & \\
               &          & (km/s)  & (km/s)   & (AU)      & \\
\hline
28 Oct. 2003   & 29 Oct.  & 2140      & 1200 & 0.993 & CME launch \\
11:00 UT       & 05:58 UT &           &      &       & to shock \\
               &          &           &      &       & arrival\\
\hline
07 Nov. 2004   & 09 Nov.  & 1190\,--\,1200 & ~800 & 0.990 & CME launch \\
15:53-16:20 UT & 02:00 UT  &           &      &       & to shock  \\
               &           &           &      &       & arrival \\
07 Nov. 2004   & 09 Nov.   & 960\,--\,970  &  &       & CME launch \\
15:53-16:20 UT & 10:00 UT  &           &      &       & to shock \\
               &           &           &      &       & arrival \\
07 Nov. 2004   & 09 Nov.   & 790\,--\,820  &  &       & to shock/\\
15:53-16:20 UT & 17:55\,--\,18:55 UT &     &  &       & magnetic cloud \\
               &           &           &      &       & arrival \\
IPS near Earth &           &           & ~700  &       & \\
\hline
20 Jan. 2005   & 21 Jan.   & 1180      & ~675  & 0.984 & flare start \\
06:36 UT       & 16:50 UT  &           &      &       & to first  \\
               &           &           &      &       & shock\\
20 Jan. 2005   & 21 Jan.   & 1130      &      &       & flare start\\
06:36 UT       & 18:20 UT  &           &      &       & to second  \\
               &           &           &      &       & shock \\
IPS near Earth &           &           & 1000 &       & \\
\hline
\end{tabular}
\end{table}

The estimated radio source heights are also in agreement with the 
CME heights on 28 October 2003. The calculated shock velocity 
from the DH type II burst at 11:16\,--\,11:21 UT (3600 km s$^{-1}$) 
indicates a shock speed only somewhat higher than the CME speed 
observed after 11:30 UT (2900 km s$^{-1}$). The white-light CME was 
also observed to decelerate. Since the high-speed full-halo CME was 
propagating in the wake of the earlier partial-halo CME, the 
high momentary speed could reflect temporary acceleration or 
abrupt changes in the local density. Estimation of the continuation 
of the DH type-II-burst lane suggests that the shock was much 
slower (2600 km s$^{-1}$) later on.

The source locations at decimetric-metric wavelengths on 28 October 
2003 have recently been analysed by Pick {\it et al.} (\citeyear{pick05}),
who discovered that the radio sources outline an H$\alpha$-Moreton 
wave, which is a direct signature of a propagating shock wave at the 
photospheric level (see {\it e.g.} Uchida, \citeyear{uchida74}). The average 
speed of the Moreton wave was around 2000 km s$^{-1}$. The authors 
also report type III bursts where the burst envelope moves with a 
speed of 2500 km s$^{-1}$, in the same direction (southwest) as the 
most compact (brightest) CME structure. Therefore it is plausible 
that the burst envelope of the type III bursts and the DH type II 
burst were closely associated with the propagating CME.     
    
In the case of 20 January 2005 we see several type-II-burst lanes in the 
dynamic spectra at decimetric-metric and DH wavelengths, and it is 
possible that several shocks were formed during the intense event. The 
shock speeds derived from the decimetric observations (650\,--\,1000 km 
s$^{-1}$) are not too far from the estimated initial CME speed, before 
acceleration (as was shown in Figure~\ref{fig4e}).

\section{Discussion on Speed Estimation}

This study of CME propagation in three different halo CME events shows
that speed estimation can be ambiguous. We list the following items 
that should be taken into account when determining the CME propagation 
characteristics: 

\begin{itemize}

\item The plane-of-the-sky component of the speed is not simple to 
deproject, supposing a radial expansion from the source region, since 
it is biased by an expansion of the CME structure that is also approaching 
us. Furthermore, the CME eruption may not necessarily be radial. In addition, 
Thomson scattering is most efficient along the perpendicular to the 
line-of-sight, which will lead to asymmetric biases in expanding structures. 
These biases are greatest in the case of frontside halo CMEs to which 
category all three events analysed here belong. Due to the biases, 
the C2 CME height constraints we used to deduce velocities using 
different density models may not give reliable results. The biases 
described also affect IPS measurements. The IPS distances to the Sun 
represent the orthogonal distances between the direction of the strongest 
disturbance and the Sun, 
which are in planes gradually turning away from the plane of the sky.

\item Simple geometric models for halo CMEs, such as the cone model, 
can give reasonable values of radial velocity provided the model 
constraints, {\it e.g.} constant velocity, are met. To solve for the two 
parameters -- cone open angle and radial velocity -- we need asymmetric 
halo CME expansion and images with good contrast, which reduces the 
number of CMEs for which the cone model may be used.

\item By making the assumption that shocks are formed at
the nose of a propagating CME (piston-driven bow shock with 
super-Alfv\'enic speed), we expect the heights of the radio shock
and the heights of the white-light CME front to match. This might
not always be the case, especially for shocks observed at 
decimetric-metric wavelengths.     

\item The heights and speeds inferred from radio burst dynamic spectra 
need to be assessed critically, since they depend strongly on the 
selected electron density model. The selection of a density model 
includes the knowledge of certain atmospheric conditions, whether 
the propagation happens in equatorial quiet-Sun densities, in dense 
coronal loops or streamer regions, in undisturbed medium, or in the 
wake of earlier transients.

\item While propagation through an undisturbed interplanetary environment 
can be traced relatively well, the task becomes more challenging when 
propagation involves interaction with slower CMEs and passage through 
a medium that has been perturbed by several significant shocks in the 
preceding days. IP scintillation observations extend the range of CME 
observation to near Earth orbit, and show in particular how the faster 
CMEs are decelerated at distances $>$60\,--\,70 {\rsun} from the Sun, 
by interaction with the solar wind plasma.   

\item {\it In-situ} measurements at the L1 point can determine the arrival 
times of shocks and ICME material and also allow estimates of the plasma
speed. However, it is not always straightforward to relate particular 
shocks and plasma clouds detected near Earth to the driving CMEs that 
left the Sun days earlier.

\end{itemize}

\begin{acknowledgements}
We thank  the Leverhulme Trust, which enabled Louise Harra (through a 
Philip Leverhulme prize award) to organise the Sun-Earth Workshop at 
Mullard Space Science Laboratory, University College London. 
J.L.C thanks the Leverhulme Trust for the award of a Leverhulme Emeritus 
Fellowship. L.v.D.G. acknowledges the Hungarian government research 
grant OTKA 048961. S.P. was partly supported by the Academy of Finland
project 104329. We thank the anonymous referee for valuable comments
and suggestions on how to improve the paper.  
We thank K-L. Klein and B. Vr\u{s}nak for discussions and help with 
the density models. 
We have used in this study radio spectral observations obtained from the 
{\it Wind} WAVES experiment 
(\texttt{http://lep694.gsfc.nasa.gov/waves/waves.html}), the 
Green Bank Solar Radio Burst Spectrometer operated by the National 
Radio Astronomy Observatory (\texttt{http://www.nrao.edu/astrores/gbsrbs/}), 
and the Hiraiso Radio Spectrograph operated by the Hiraiso Solar 
Observatory, National Institute of Information and Communications 
Technology (\texttt{http://sunbase.nict. go.jp/solar/denpa/}). 
We thank their staff for making the data available at their Web 
archives, and we thank K. Hori for preparing the HiRAS spectral plot.  
The Ooty Radio Telescope is operated by National Centre for Radio 
Astrophysics, Tata Institute of Fundamental Research. 
We are grateful to the SOHO LASCO and SOHO EIT teams for making
their data available in their Web archives. SOHO is a project of 
international cooperation between ESA and NASA.
The LASCO CME Catalog is generated and maintained by NASA and Catholic
University of America in co-operation with Naval Research Laboratory. The
catalogue can be accessed at \texttt{http://cdaw.gsfc.nasa.gov/CME\_list/}.
\end{acknowledgements}

\end{article} 
\end{document}